\pdfoutput=1

\documentclass[usenatbib,usegraphicx]{mn2e}
\usepackage{amssymb,bm,color,charter}
\usepackage{ulem}

\newcommand{\e}[1]{\times 10^{#1}}                        % x 10^...
\newcommand{\fig}[1]{Fig. \ref{#1}}

\defcitealias{LambEtAl08a}{a}
\defcitealias{LambEtAl08b}{b}

\newcommand{\fignar}[4][1]{
   \begin{figure}
   \centering
   \includegraphics[scale=#1]{#2}
   \caption{#4}
   \label{#3}
   \end{figure}
}

\newcommand{\figwide}[4][1]{
   \begin{figure*}
   \centering
   \includegraphics[scale=#1]{#2}
   \caption{#4}
   \label{#3}
   \end{figure*}
}
\title{Discovery of Drifting High-frequency QPOs in Global Simulations
of Magnetic Boundary Layers}

\author[M. M. Romanova \& A. K. Kulkarni]{
M. M. Romanova\thanks{E-mail: romanova@astro.cornell.edu},
A. K. Kulkarni\thanks{E-mail: akshay@astro.cornell.edu}\\
Dept. of Astronomy, Cornell University, Ithaca, NY 14853}

\begin{document}

\maketitle

\begin{abstract}

\noindent We report on the numerical discovery of quasi-periodic oscillations
(QPOs) associated with accretion through a non-axisymmetric magnetic
boundary layer in the unstable regime, when two ordered equatorial streams form and rotate
synchronously at approximately the angular velocity of the inner disk
The streams hit the star's surface
producing hot spots. Rotation of the spots leads to high-frequency QPOs.
We performed a number of simulation runs
for different magnetospheric sizes from small to tiny, and observed a definite correlation
between the inner disk radius and the QPO frequency: the frequency is higher
when the magnetosphere is smaller.
In the stable regime a small magnetosphere forms and accretion through the
usual funnel streams is observed, and the frequency of the star
is expected to dominate the lightcurve. We performed exploratory investigations
of the case in which the magnetosphere becomes negligibly small
and the disk interacts with the star through an equatorial belt. We also performed investigation
of somewhat larger magnetospheres where one or two ordered tongues may dominate over other chaotic tongues.
In application to millisecond pulsars we obtain
QPO frequencies in the range of 350 Hz to 990 Hz
for one spot. The frequency associated with rotation of one spot may dominate if
spots are not identical or antipodal. If the spots are similar and antipodal
then the frequencies are twice as high. We show that variation of the accretion
rate leads to drift of the QPO peak.

\end{abstract}


%=========================================================

\section{Introduction}

In disk-accreting magnetized stars, disk-magnetosphere
interaction
and channeling of matter to the magnetic poles may determine the
majority of observational features of such stars
\citep[e.g.,][]{GhoshLamb79, Konigl91}. These include young solar-type stars
\citep[e.g.,][]{BouvierEtAl07} and very old stars such as magnetized white
dwarfs \citep{Warner95} or accreting millisecond pulsars \citep{vanderKlis00}.
In many cases no clear period is observed, and high-frequency
quasi-periodic oscillations (QPOs) with a frequency corresponding to
a fraction of the Keplerian frequency at the stellar surface are
seen instead. It is often the case that a second, lower frequency
peak appears and drifts in concert with the main frequency, forming
two-peak QPOs. Such high-frequency oscillations are observed in both
accreting millisecond pulsars \citep[e.g.,][]{WijnandsvanderKlis98}
and in dwarf novae, which are a sub-class of cataclysmic variables
\citep{WarnerWoudt02a, PretoriusWarnerWoudt06}. It has
been suggested that the high-frequency QPOs may be associated with
accretion to a star with a very weak magnetic field, when some sort
of weakly magnetized equatorial belt forms \citep{Paczynski78, WarnerWoudt02b}.
\citet{MillerLambPsaltis98} and \citet{LambMiller01} suggested that the
high-frequency QPOs in accreting millisecond pulsars may result from
rotation of clumps in the inner disk around a weakly magnetized
star. They suggested that some of the disk matter may penetrate
through the magnetosphere until stopped, e.g., by the radiation pressure.
In this paper we investigate accretion to stars with {\it small
magnetospheres}, where the energy density of the disk matter
dominates up to very small distances from the star, or even up to
the stellar surface, while magnetospheres are small compared to the
radius of the star. We call the region around the star with a small magnetosphere
 the Magnetic Boundary Layer (MBL), and consider
different cases. We notice that as for large magnetospheres, matter
may accrete either in the stable (e.g., \citealt{RomanovaEtAl03, RomanovaEtAl04}, hereafter RUKL03
and RUKL04) or
unstable regime (\citealt{KulkarniRomanova08a, KulkarniRomanova08b}, hereafter KR08a,b;
\citealt{RomanovaKulkarniLovelace08}, hereafter RKL08). The most
striking result is the fact that in the unstable regime, an ordered
structure of two tongues forms, and rotates with the angular velocity of
the inner disk, showing clear high-frequency QPO peaks associated
with this rotation\footnote{See animations at http://wwww.astro.cornell.edu/$\sim$romanova/qpo.htm}.
Unstable accretion with a preferred number
of tongues as a possible origin for QPOs has been discussed earlier by \citet{LiNarayan04}.
In the present research we have shown that
an even more ordered situation is possible
with small magnetospheres, where
two ordered tongues rotate with the frequency of the inner disk.
Variation of the accretion rate leads to drifting of this QPO frequency.
Correlation between the frequency and accretion rate has been observed in a number of
accreting millisecond pulsars \citep[e.g.,][]{vanderKlis00}. Below we describe this and other
results in detail. In \S2 we discuss a possible classification scheme for
magnetic boundary layers. In \S 3 we describe our model and reference parameters.
In \S4 we consider MBLs during stable
accretion. In \S5 we describe MBLs and formation of QPOs in the
unstable regime. We present some discussion and conclusions in \S6.

%=========================================================

\section{Classification of boundary layers}

In accreting magnetized stars the disk is stopped by the
stellar magnetosphere at the magnetospheric, or truncation,
radius \citep[e.g.,][]{ElsnerLamb77}:
$$
r_m=k_A({GM_*})^{-1/7}\dot{M}^{-2/7}\mu^{4/7},
$$
where  $M_*$ and $\mu$ are the mass and magnetic moment of the star,
$\dot M$ is the accretion rate through the disk, and the coefficient
$k_A$ is of the order of unity \citep[e.g.,][]{LongRomanovaLovelace05, BessolazEtAl08}.
The magnetospheric radius may be small
if the magnetic moment of the star $\mu$ is small, or if the
accretion rate is high. At a given magnetic moment of the star,
variation of the accretion rate will lead to expansion or
compression of the magnetosphere, and so at very high accretion
rates the magnetosphere may be completely buried by accreting matter
\citep[e.g.,][]{CummingZweibelBildsten01, LovelaceRomanovaBisnovatyiKogan05},
while at very low accretion rates it
strongly expands, and different types of accretion are expected at
different $\dot M$ \citep[e.g.,][]{RomanovaEtAl08}. An important
parameter of the problem is the size of the magnetosphere in units
of the stellar radius, $r_m/R_*$. Depending on this parameter we can
define three types of boundary layers:
\smallskip

\noindent{\it (1).~ Hydrodynamic BL}. If the ratio $r_m/R_* << 1$,
then the magnetic field does not influence the dynamics of matter
flow in the vicinity of the star, and a purely hydrodynamic boundary
layer is expected where the magnetic field can be neglected. In this
case the disk matter comes to the surface of the star and
interacts with the star in the equatorial zone
\citep[e.g.,][]{LyndenBellPringle74, Pringle81, PophamNarayan95}. Such interaction leads
to release of a significant amount of energy at the surface of the star, about a
half of the gravitational energy, $\dot L_{BL}\approx 0.5 GM_* \dot M/R_*$.
Interaction through such a hydrodynamic boundary layer is expected in many
non-magnetic white dwarfs and neutron stars and in cases where
accretion rate is so high that the magnetic field is buried by the accreting
matter. It has been noted that the boundary layer matter
will not interact with the star as a thin layer, but
will spread to higher latitudes along the surface of the star due
to matter pressure \citep{FerlandEtAl82}. This effect has been calculated
by \citet{InogamovSunyaev99} and \citet{PiroBildsten04} and has
recently been observed in axisymmetric hydrodynamic simulations by
\citet{FiskerBalsara05}. Much less work has been done for analysis
of the magnetic boundary layers (MBL).

\smallskip

\noindent{\it (2).~ Magnetic Boundary Layer (MBL) With
Magnetospheric Gap}. If the magnetosphere is only slightly larger
than the radius of the star, $r_m/R_* > 1$, then it stops the disk
at a small distance from the star. The simulations described in this
paper show that in this case, accretion may be either in the stable
regime producing funnel streams, or in the unstable regime where most of the matter accretes
through instabilities (KR08a; RKL08). We find that an unusual type of
MBL forms in the unstable regime with two ordered streams rotating
in the equatorial plane at approximately the angular velocity of the
disk. These streams may be responsible for high-frequency QPOs.

\smallskip

\noindent{\it (3).~MBL Without a Magnetospheric Gap}. If the
magnetospheric radius is comparable to the stellar radius,
$r_m/R_*\approx 1$, then the magnetosphere cannot stop the disk, and the
disk interacts with the stellar surface. In this case the magnetic field
may be somewhat enhanced in the accretion disk due to wrapping of
the magnetic field lines by the disk matter. Preliminary simulations
have shown that interesting phenomena may happen at the
disk-magnetosphere boundary, such as interaction through
a Kelvin-Helmholtz-type instability.

\smallskip
Global 3D simulations of accretion to stars with large
magnetospheres (a few stellar radii in size) have been performed earlier
(e.g., RUKL03,04; KR05).
It was recently found that stars may be either in
the stable or unstable regime of accretion (KR08a,b; RKL08). In this
paper we show results of global 3D simulations of accretion to star
with very small magnetospheres, in both stable and unstable regimes,
and some results for direct interaction between the disk and the star.

%=========================================================

\section{Model}

We solve the 3D MHD equations presented, e.g., in RUKL03, using the
``cubed sphere" second order Godunov-type code described in
\citet{KoldobaEtAl02}. The equations are written in a coordinate system rotating with a star.
The energy equation is solved for the entropy, and the equation of state is that for an ideal gas.
Viscosity is incorporated into the code with a viscosity coefficient proportional to $\alpha$
\citep{ShakuraSunyaev73}, where $\alpha < 1$. The action of the viscosity is restricted to
regions of high density (the disk). Viscosity helps  bring matter towards the star from the disk.

\subsection{Dimensionalization and Reference Values}

The MHD equations are solved in
dimensionless form so that the results can be readily applied to different types of stars.
We have the freedom to choose three parameters from which the reference scales are derived,
and we choose those to be the star's mass $M$, radius $R_*$ and magnetic moment
$\mu_*$ (with corresponding magnetic field $B_*=\mu_*/R_*^3$).
We take the reference
mass $M_0$ to be the mass $M$ of the star.
The reference radius is about three times the radius of the star,
$R_0=R_*/0.35\approx 3\times R_*$. In our past research (RUKL03; RUKL04) this radius approximately
corresponded to the truncation (or magnetospheric) radius. In the present research the truncation radius
is much smaller, but we keep this unit for consistency with our prior work.
The reference velocity is $v_0=(GM/R_0)^{1/2}$.
The reference time-scale $t_0=R_0/v_0$, and the reference angular velocity $\omega_0=1/t_0$.
We measure time in units of $P_0=2\pi t_0$ (which is the Keplerian rotation period at $r=R_0$).
In the plots we use the dimensionless time $T=t/P_0$.

The reference magnetic moment  $\mu_0=\mu_*/\widetilde{\mu}$, where $\widetilde{\mu}$
is a dimensionless parameter which determines the dimensionless size of the magnetosphere.
The reference magnetic field is $B_0=\mu_0/R_0^3$.
The reference density is $\rho_0 = B_0^{2}/v_0^{2}$. The reference accretion rate is
$\dot M_0 = \rho_0 v_0 R_0^{2}$.  Taking into account relationships
for $\rho_0$, $B_0$, $v_0$ we obtain $\dot M_0=\mu_*^2/(\widetilde{\mu}^2 v_0 R_0^4)$.
The dimensional accretion rate
is $\dot M=\widetilde{\dot M}\dot M_0$, where
$\widetilde{\dot M}$ is the dimensionless accretion rate onto the surface of the star which is obtained
from the simulations. Substituting parameters of the star to the reference values, we obtain the
dimensional accretion rate:
$$
\dot M = \bigg(\frac{\widetilde{\dot M}}{{\widetilde\mu}^2}\bigg)\frac{\mu_*^2}{(GM)^{1/2} (R_*/0.35)^{7/2}}.
\eqno(1)
$$
One can see that at fixed parameters of the star ($M, R_*, \mu_*$), the accretion rate depends on the ratio
$\widetilde{\dot M}/{\widetilde{\mu}}^2$. Further analysis shows that
parameter ${\widetilde{\mu}}^2$ varies more strongly
than $\widetilde{\dot M}$ and determines the
variation of the accretion rate. In the paper we vary parameter $\widetilde{\mu}$ which
leads to variation of the dimensional accretion rate.
Table \ref{tab:refval} shows examples of reference variables for different stars.
We solve the MHD equations using the normalized variables
$\widetilde\rho=\rho/\rho_0$, $\widetilde v=v/v_0$, $\widetilde B= B/B_0$, etc.
Most of the plots show the normalized variables (with the tildes implicit). To
obtain dimensional values one needs to multiply values from the plots by the
corresponding  reference values from Table \ref{tab:refval}.

\subsection{Initial and Boundary Conditions}

Here we briefly summarize the initial and boundary conditions,
described in detail in our earlier papers (e.g. in RUKL03).
The simulation region consists of the accretion disk, the corona and the star. The star has a dipole
magnetic field which is frozen into its surface.
The disk is cold and dense. The corona is 100 times hotter and
100 times less dense (at the fiducial point at the inner edge of the disk). Initially
we rotate the disk with Keplerian velocity and the star with the angular velocity
of the inner disk,
so as to avoid discontinuity of the magnetic field lines at the disk-star boundary.
We also rotate the corona above the disk with the Keplerian velocity of the disk so as
 to avoid discontinuity at the disk-corona boundary. This is necessary since
  the discontinuity may lead to artificially strong forces
on the disk and strong magnetic braking. In addition, we derive such a distribution of the density
and pressure in the disk and corona, that the sum of the gravitational, pressure and centrifugal forces
is equal to zero everywhere in the simulation region \citep[]{RomanovaEtAl02}.
This initial condition ensures a very smooth gradual start-up, where the disk matter accretes slowly inwards
due to viscous stresses. On the other hand, this condition dictates a density distribution that
does not correspond to that in an
equilibrium viscous disk \citep[e.g.][]{ShakuraSunyaev73}. That is why our disk starts reconstructing
itself on the viscous time-scale. Fortunately,  matter from the inner parts
of the disk flows inward for a long time, and the matter
flux increases with $\alpha$, though dependence is not linear (as one would expect from the theoretical
prediction for viscous disks). In addition, interaction of the disk with the external magnetosphere
leads to some magnetic
braking, leading to an enhancement of the accretion rate which may be smaller or larger than the effect
of the viscosity
and also to oscillations of the matter flux. That is why we cannot
use the dimensionless accretion
rate $\widetilde{\dot M}$ as a parameter of the problem. Instead, we choose
 $\widetilde{\mu}$ as the main
parameter which is responsible for the size of the magnetosphere. Equation (1) shows that the dimensional
accretion rate onto the star is regulated by the combination $\widetilde{\dot M}/{\widetilde\mu}^2$.
In this paper we consider small magnetospheres with
$\widetilde{\mu}\sim 0.08-0.5$, which is
about 10 times smaller than in our past research (RUKL03; RUKL04; KR05).

The size of the simulation region is $R_{max}=15 R_0\approx 45 R_*$.
We initially place the inner radius of the disk at a distance $r_d$ from the star
which is either $1.2 R_0\approx 3.6 R_*$ (in stable cases)
or $1.8 R_0\approx 4.8 R_*$ (in the unstable cases). Simulations
show that the result does not depend on $r_d$, because in both cases the disk matter
moves inward and is stopped by the magnetosphere at small distances from the star.
Then we have the freedom to spin the star up or down gradually. In this paper we choose
a slowly rotating star with angular velocity $\omega_*=(GM/r_{cor})^{1/2}$ corresponding to
the corotation radius $r_{cor}=2 R_0$. In dimensionless units, $\widetilde{\omega_*}=0.354$, and in
application to neutron stars, $P_*=6.2$ms.

\smallskip
{\bf The boundary conditions}  are similar to those used in our other papers (e.g., RUKL03).
On the star (the inner boundary) the magnetic field is frozen into the surface of the star
(the normal component of the field, $B_n$, is fixed), though all three
magnetic field components may vary.
For all other variables $A$, free conditions are prescribed: $\partial A/\partial n = 0$.
The total velocity vector is adjusted to be parallel to the total magnetic field
to enhance the frozen-in condition. At the external boundary, $r=R_{out}$, again free boundary
conditions are prescribed for all variables. In some cases we fix the density
on the external boundary
so that matter of the disk stays in the region during the
 viscous reconstruction mentioned above.
We do not supply matter to the disk from the external boundary, because
the amount of matter in the inner regions of the disk is sufficient to support accretion
over the duration of our simulations.

\smallskip
{\bf The grid.} Simulations of accretion
to stars with small magnetospheres require a finer grid compared to the larger
magnetosphere cases (e.g. RUKL03).
We use a grid resolution of $N_r\times N^2 = 100\times 41^2$ (in each of the 6 blocks of the
cubed sphere) for most of our cases, and $N_r\times N^2 = 130\times 61^2$
for the smallest magnetospheres.
The simulations were done using 60-180 processors each on the NASA HPC clusters.

\section{Accretion in the stable regime}

Below we describe the results of simulations of stable accretion
to stars with small magnetospheres $\widetilde{\mu}=0.08 - 0.2$.

\begin{table*}

%\begin{tabular}{l@{\extracolsep{0.2em}}l@{}lll}
\begin{tabular}{llll}
%_____________________________________________________________________________________________________________________
\hline                & CTTSs              & White dwarfs   & Neutron stars      \\
%__________________________________________________________________________________
\hline
$M(M_\odot)$          & 0.8                & 1              & 1.4                \\
$R_*$                   & $2R_\odot$         & 5000 km        & 10 km              \\
$B_*$ (G)     & $10^3$             & $10^6$         & $3\e8$             \\
$R_0$ (cm)            & $4\e{11}$          & $1.4\e9$       & $2.9\e6$           \\
 $v_0$ (cm s$^{-1}$)   & $1.6\e7$           & $3\e8$         & $8.1\e9$           \\
$\omega_0$ (s$^{-1}$) & $4\e{-5}$          & 0.21           & $2.8\e3$           \\
$\nu_0$               & $0.55$ day$^{-1}$  & $3.2\e{-2}$ Hz & $4.5\e2$ Hz        \\
$P_0$                 & $1.8$ days         & 29 s           & 2.2 ms             \\
%$\mu_*$ (Gcm$^3$)       & $2.7\e{36}$          & $1.2\e{32}$      & $1.0\e{27}$         \\
%$\dot M_0$ ($M_\odot$yr$^{-1}$) & $2.8\e{-7}$ & $1.9\e{-7}$ & $6.5\e{-8}$        \\
%__________________________________________________________________________________
\hline
\end{tabular}

\caption{Sample reference values of the dynamical quantities for
different types of stars. We choose the mass, radius and magnetic field
of the star, and define the other variables in terms of these three quantities.
Note that the frequency $\nu_0$ is the Keplerian frequency at radius $R_0$.} \label{tab:refval}

\end{table*}

Simulations show that the boundary between the stable and
unstable regimes depends on a number of factors, such as the accretion rate
(determined by the viscosity parameter $\alpha$), the rotation period
 of the star $P_*$, and the misalignment
angle $\Theta$ of the dipole (KR08a,b). If a star with a small misalignment angle,
$\Theta < 5^\circ$, rotates slowly (the disk rotates much
faster than the star), then the accretion has a tendency to be unstable
irrespective of $\alpha$ (KR08a,b).
We have chosen slowly rotating stars with dimensionless angular velocity
 $widetilde{\omega_*}=0.354$ (corresponding to a corotation radius of $r_{cor}=2$). We take
slowly rotating stars in order to model the situation in which the
star initially has a large magnetosphere and is in the rotational
equilibrium state, but later the accretion rate increases and the
magnetosphere becomes small. To stabilize accretion we chose
$\Theta=15^\circ$. We also used $\alpha=0.04$. In the description of
the results we measure time in units of the Keplerian rotation
period at $r=1$. We consider two main cases, one with
$\widetilde{\mu}=0.2$ which has a small magnetosphere, and the other with
$\widetilde{\mu}=0.08$ which has a tiny magnetosphere.

%%%%%%%%%%%%%%%%%%%%%%%%%% Fig 1 %%%%%%%%%%%%%%%%%%%%%%%%%%

\fignar{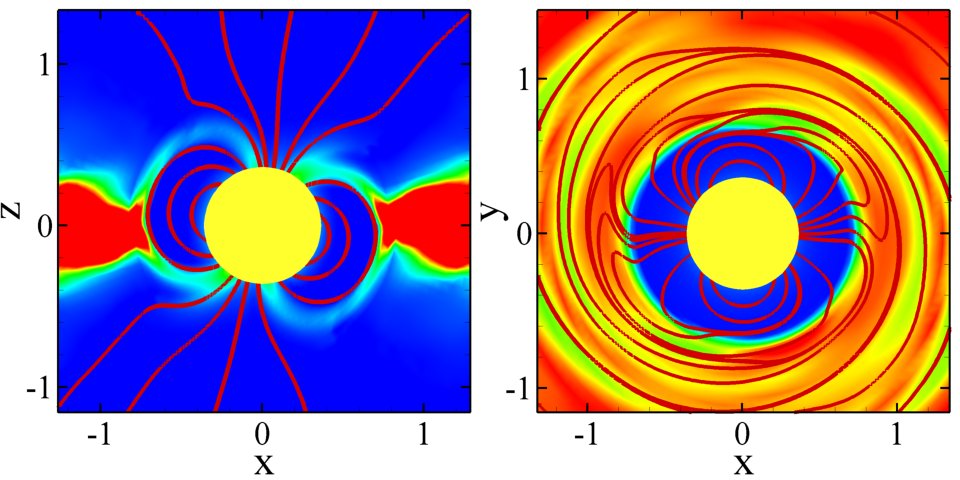}{fun-small-2}{Stable accretion to a star with a small magnetosphere,
where the magnetic moment of the star is $\widetilde{\mu}=0.2$ and is misaligned
relative to the rotational axis $\Omega$ at $\Theta=15^\circ$. {\it Left
panel:} Density distribution and sample magnetic
field lines in the $xz$ ($\mu-\Omega$) plane. {\it Right panel:} The same, but in the $xy-$
(equatorial) plane. The dimensionless density varies between 0.8 (red) and
0.006 (blue).}

%%%%%%%%%%%%%%%%%%%%%%%%%%%%%%%%%%%%%%%%%%%%%%%%%%%%%%%%%%%

\fig{fun-small-2} shows an example of  accretion to a star with
$\widetilde{\mu}=0.2$. A small
magnetosphere forms, in which the magnetic field of the dipole
dominates. The position of the inner disk edge is determined by the
balance between the magnetic and kinetic matter pressure. Matter is
lifted above the magnetosphere forming two funnel streams, and
accretes to the surface of the star forming two hot spots. The hot spots
show the distribution of the total energy flux on the surface of the star
(see RUKL04 and KR05 for details).

The hot spots have a preferred position, although in case of small misalignment angles
the funnel streams are dragged by the rapidly rotating disk and may rotate faster than the star.
Such spot rotation has been observed in earlier simulations
(RUKL03; \citealt{RomanovaEtAl06}). In the opposite situation when the star spins fast,
both funnels and spots may rotate
slower than the star (e.g. Romanova et al. 2002). Faster or slower rotation
of spots may lead to a QPO feature with
frequency higher or lower than the stellar frequency (Romanova et al. 2006).
It has been suggested that at small misalignment angles, the spots
may wander around the magnetic poles, possibly
causing intermittency in some millisecond pulsars
% \citeauthor{LambEtAl08a} (2008a) and \citealt{LambEtAl08b}
\citeauthor{LambEtAl08a} (\citeyear{LambEtAl08a}\citetalias{LambEtAl08a},\citetalias{LambEtAl08b})
(see also \citep{AltamiranoEtAl08, CasellaEtAl08}). On the
other hand, if the misalignment angle of the dipole is not very small,
then such faster or slower spot motions become less significant. The spots
acquire a preferred place on the surface of the star and corotate with the star
(see also RUKL03; RUKL04; KR05). In simulations
with $\Theta=15^\circ$ we observe a mixture of spots: on the one hand, the disk
drags the funnel around the weak magnetosphere in spite of the relatively high
misalignment angle, and frame to frame analysis shows that the spots often move
faster than the star.
On the other hand, the hot spots spend a somewhat longer time in the $\mu-\Omega$ plane
compared with other positions, due to which we observe a definite peak
associated with star's spin in the Fourier spectrum. Longer simulation runs
are probably required to see the QPO peak.

%%%%%%%%%%%%%%%%%%%%%%%% Fig 2 %%%%%%%%%%%%%%%%%%%%%%%%%%%%%%%

\figwide{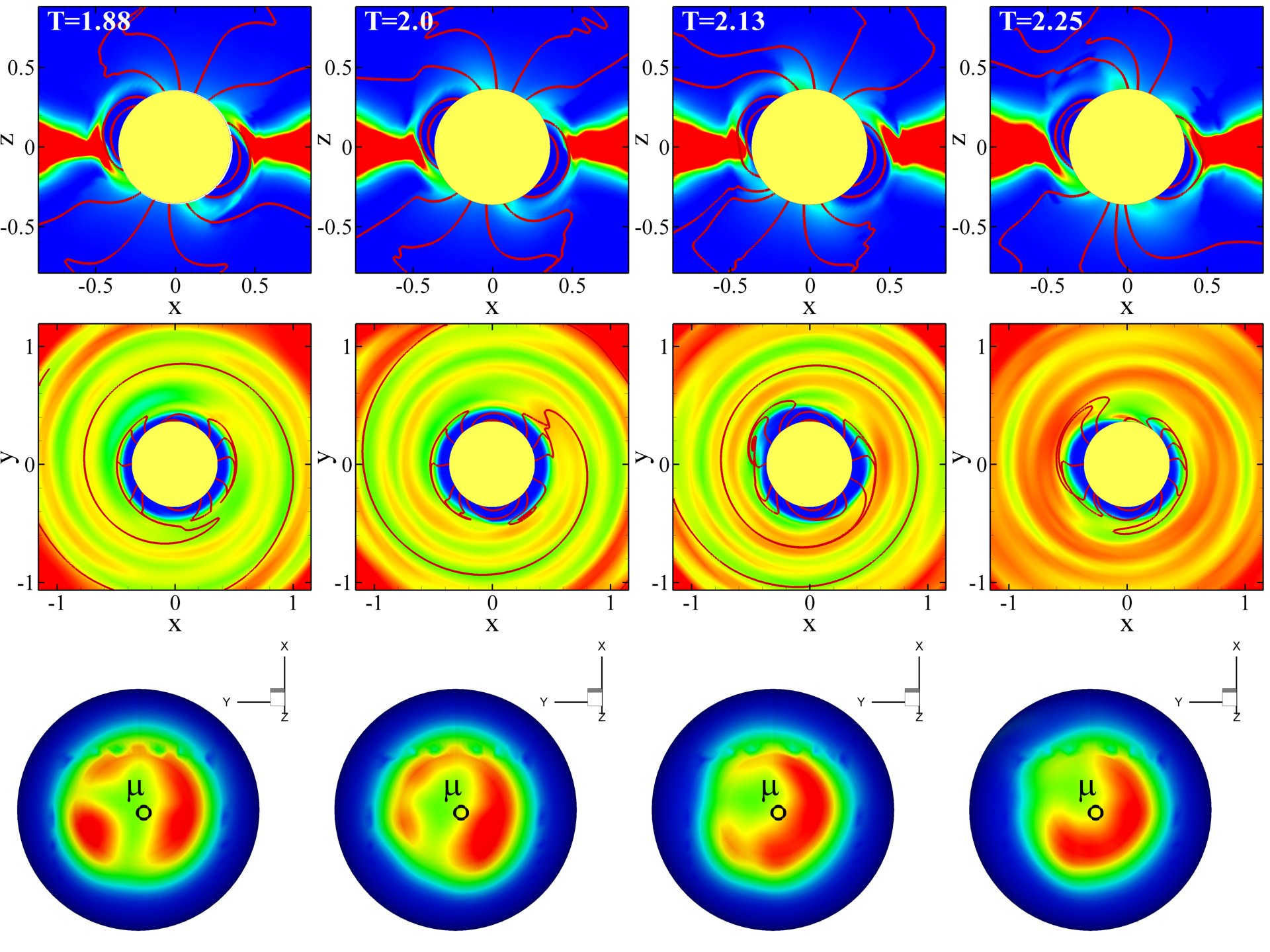}{fun-tiny-12}{Stable accretion to a star with a
very small (tiny) magnetosphere ($\widetilde{\mu}=0.08$) with
$\Theta=15^\circ$ ($\alpha=0.04$). {\it Top two panels:}
Density distribution and sample magnetic field lines in the $xz$
($\mu-\Omega$) plane and the $xy-$ (equatorial) plane in a
coordinate system rotating with the star. The dimensionless density
varies between 1.4 (red) and
0.006 (blue). {\it Bottom panel:} Distribution of the dimensionless energy flux of matter onto the
surface of the star (pole-on), varying between 0.4 (red)
and 0.0009 (blue).  The time $T$ is measured in units of $\Delta
T=1/16 P_0$, where $P_0$ is the Keplerian period at $r=1$.}

\fig{fun-tiny-12} shows an example of accretion to a star with
a {\it tiny magnetosphere}, $\widetilde{\mu}=0.08$. It is amazing that
in this case too, the magnetosphere channels the accreting matter, forming tiny funnel
streams hitting the surface of the star. Density waves form in the
equatorial plane of the disk. Sometimes density fluctuations in the
accreting matter push the disk to the stellar surface. However,
later the magnetosphere is restored again. The hot spots have a
preferred position but they often rotate faster than the star as a
result of the difference in angular velocities between the disk and
the star. In this case again, a QPO at the stellar frequency is
expected, and also a QPO corresponding to the frequency at the inner
edge of the disk and beats between these two frequencies. Longer
simulation runs are required to extract these frequencies.

The main result of this section is that even small and tiny
magnetospheres disrupt the disk and channel matter to the star,
forming hot spots. It is also important to note that the
high-frequency QPO associated with rotation of the inner edge of the
disk is expected if the disk rotates faster (slower) than the star.
Signs of faster rotation of the spot are clearly observed. Drifting of the
high-frequency QPO is expected if the inner edge of the disk
varies as a result of variation of the accretion rate. Future
(longer) simulations will help obtain different frequencies.

%%%%%%%%%%%%%%%%%%%%%%%% Fig 3 %%%%%%%%%%%%%%%%%%%%%%%%%%%%%%%

\figwide{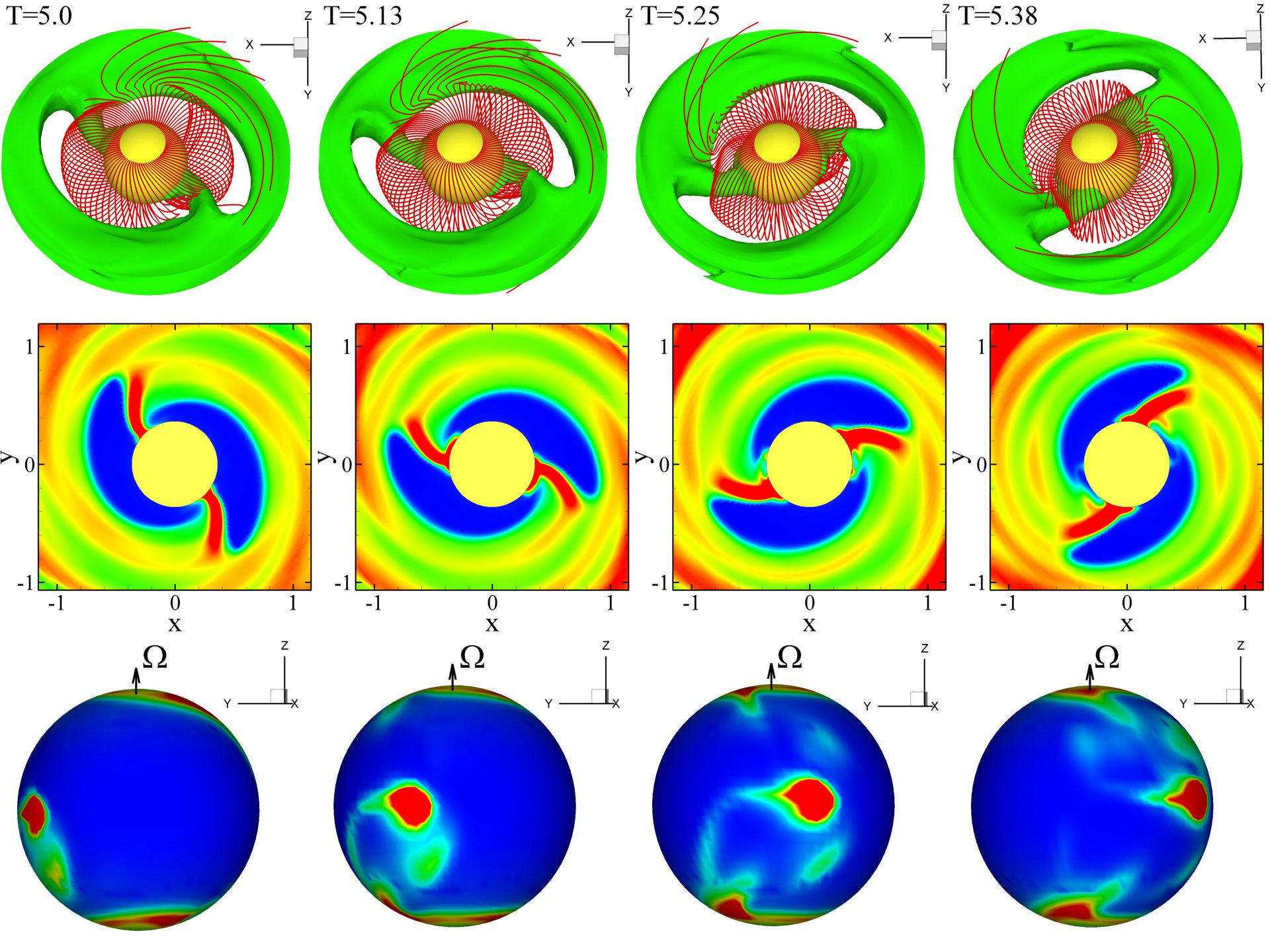}{mbl-small-12}{Snapshots of MBL accretion
through the synchronized
instability tongues in the case of a relatively large magnetosphere
($\widetilde{\mu}=0.2$ in dimensionless units). The time interval shown is a small
portion of the simulation;
the tongue pattern is steady for the entire duration of the simulation.
{\it Top panel:} A surface of constant
density (0.4 in dimensionless units) and sample magnetic field lines.
{\it Middle panel:} Density
distribution and sample magnetic field lines in the $xy$-plane.
The density varies from 6.7 (red) to 0.01 (blue).
{\it Bottom panel:} Energy flux onto the star's surface, ranging from 3.4
(red) to 0.008 (blue).
The figures are shown in a coordinate system rotating with the star. The time $T$ is
measured in periods of Keplerian rotation at $r=1$.}

%%%%%%%%%%%%%%%%%%%%%%%% Fig 4 %%%%%%%%%%%%%%%%%%%%%%%%%%%%%%%

\figwide{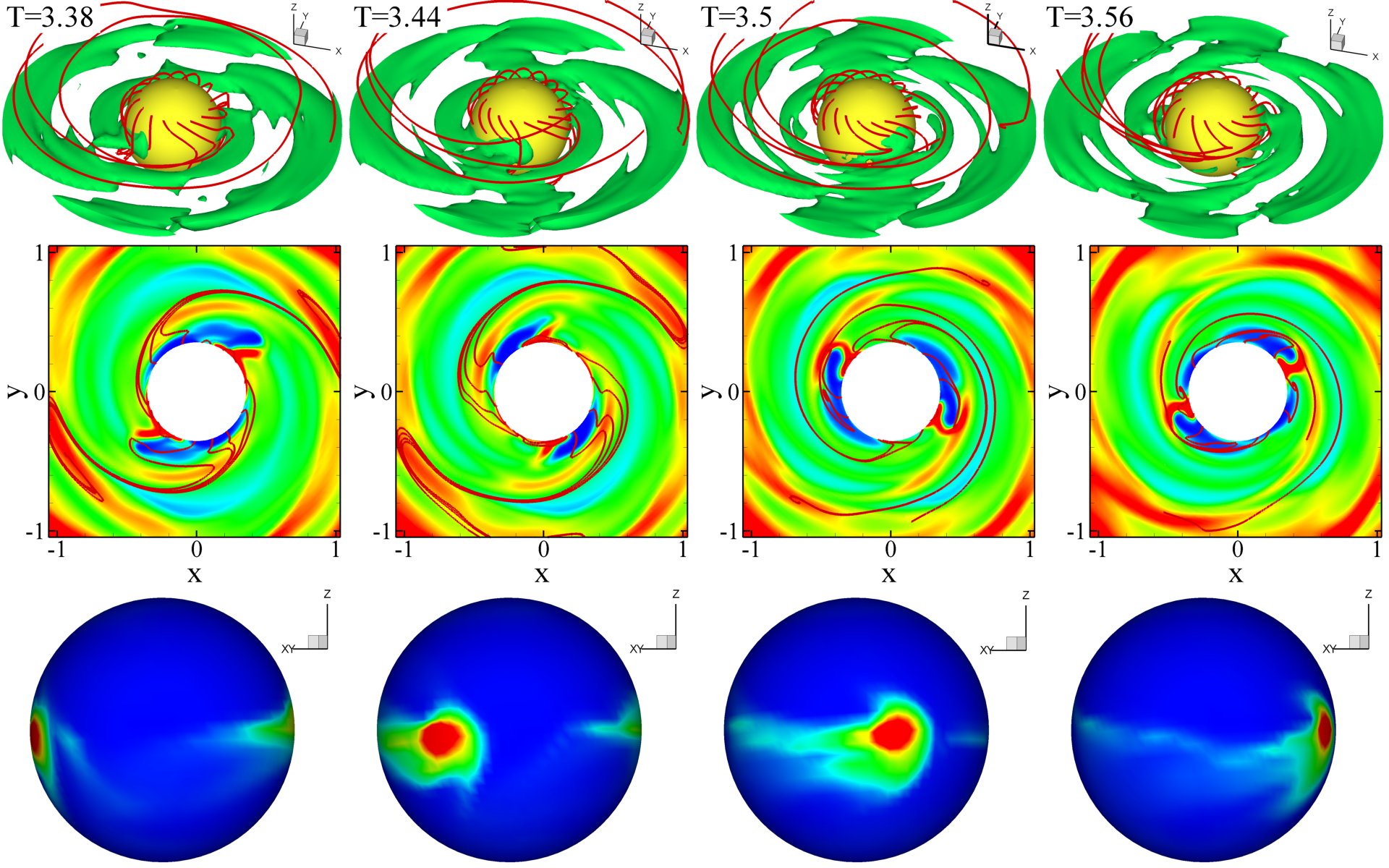}{mbl-tiny-12}{{\it Top panel:} A surface of constant
density (0.47 in dimensionless units) and sample magnetic field lines
in the case of a small magnetosphere, $\widetilde{\mu}=0.1$. {\it Middle panel:} $xy$-slice showing the
density distribution and sample magnetic field lines. {\it Bottom panel:}
density distribution on the surface of the star, ranging from 3.8 (red)
to 0.01 (blue). The figures are
shown in a coordinate system rotating with the star. The time $T$ is
measured in periods of Keplerian rotation at $r=1$.}

%%%%%%%%%%%%%%%%%%%%%%%% Fig 5 %%%%%%%%%%%%%%%%%%%%%%%%%%%%%%%

\figwide{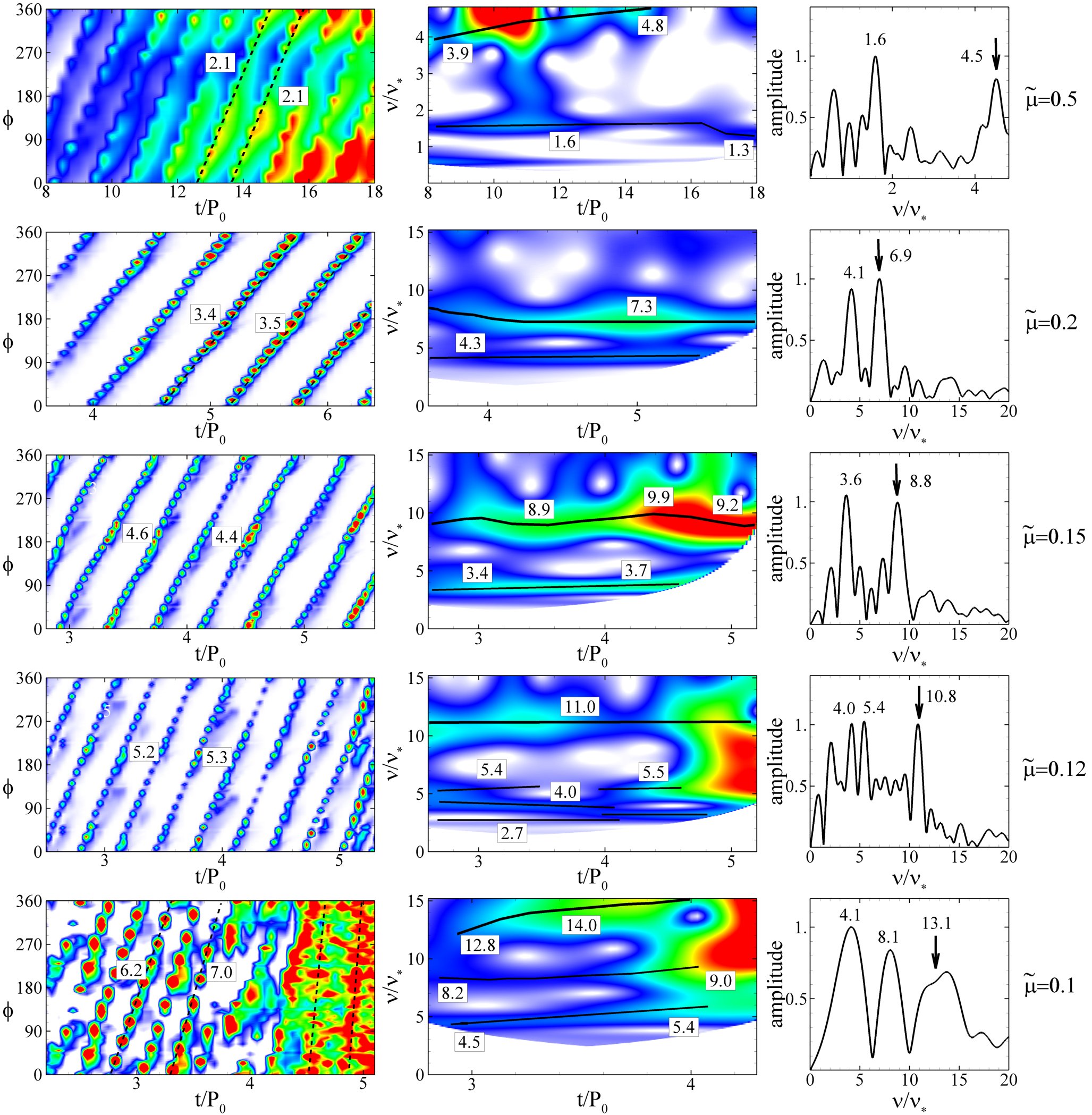}{swf-all}{Frequency analysis for runs with different
magnetospheric sizes. {\it Left column:} Spot-omega analysis showing
the emitted energy flux distribution in the equatorial
plane at different times. The flux varies from 3.4 (red) to 0.008
(blue). The lines reflect the motion of individual spots on
the stellar surface. The slope of the lines is proportional to the
angular velocity of the hot spots.
{\it Right two columns:} Wavelet and Fourier spectra of the light
curves from the hot spots
obtained at an observer inclination angle of $i=90^\circ$ (i.e., when
the observer is in the
equatorial plane). The arrows show the frequency corresponding to
the presence of two rotating spots.
In the bottom row the lines show the distribution of the density, and
not the flux, in the spots, and the density varies between
3.8 (red) and 0.01 (blue). In this case the lightcurve
for which the wavelet and Fourier
spectra are shown is calculated assuming that the emission of the star
is proportional to the density.}

%=========================================================

\section{Accretion in the unstable regime: A New Type of Boundary Layer and QPOs}

\subsection{Formation of two symmetric streams and spots}

Now we take a star with a small misalignment angle, $\Theta=5^\circ$, and investigate accretion
in the unstable regime. First we choose a small magnetosphere with magnetic moment $\widetilde{\mu}=0.2$.
 We  take the viscosity coefficient $\alpha=0.1$ in all runs below. A star rotates slowly
 with angular velocity $\widetilde{\omega_*}=0.354$ (corresponding to $r_{cor}=2$).
  We observe that the instability starts,
and matter penetrates through the equatorial region of the magnetosphere
through a number of tongues. Later, however, two symmetric ordered tongues form
and rotate synchronously with angular velocity approximately
equal to the angular velocity in the inner region of the accretion
disk, that is,  much faster than the star. These tongues reach the
surface of the star and produce two hot spots at  the star's equator
which move faster than the star.
The tongues deposit a significant amount of energy to the surface of the star. Part of
this is the gravitational energy associated with acceleration of
matter towards the star. We take this energy into account while
plotting hot spots. In addition, there is energy released due to friction
between the surface of the slowly rotating star and the foot-points of
the rapidly rotating tongue.

\fig{mbl-small-12} shows snapshots of rotation of the
tongues, slices in the equatorial plane and hot spots (energy flux) on the star's
surface. The slices show that the magnetosphere has a strongly
modified shape, particularly in those places where the tongues reach
the surface of the star. They push the magnetosphere to the surface
of the star and when the tongue moves, the magnetosphere re-emerges,
while the tongue pushes another part of the magnetosphere to the
surface of the star. Two strong hot spots form close to equatorial
region. Sometimes they are not symmetric because the tongues push
the magnetosphere more to one side than another. Some matter
accretes to the poles through weak funnel streams. It is interesting
that the polar spots also rotate with the angular velocity of the
tongues. Thus, in the case of a small magnetosphere we observe {\it
a new phenomenon - a modified boundary layer}, where the funnel
streams and hot spots move with the angular velocity of the disk ---
much faster than the star. This is a new type of boundary layer
where matter of the disk interacts with the star through unusual
symmetric tongues.  Rotation of the hot spots along the surface of the
star is expected to give strong QPO peaks at twice the frequency of the inner disk.

To check this phenomenon, we performed a set of simulation runs
at a variety of stellar magnetic moments: $\widetilde{\mu}=0.16$, $\widetilde{\mu}=0.15$,
$\widetilde{\mu}=0.12$, $\widetilde{\mu}=0.1$,
keeping all the other parameters fixed. We observed that
similar symmetric streams form and rotate around the star.
However, at smaller $\widetilde{\mu}$ the disk stops closer to the star.
In the case of $\widetilde{\mu}=0.1$, only a tiny magnetosphere forms
with tiny streams (see \fig{mbl-tiny-12}).
The bottom panels of the \fig{mbl-tiny-12} show the
density distribution in the hot spots instead of the emitted flux,
because it is expected that energy would be released mainly due to
friction between the tongues and the surface of the star, which we
do not calculate in this paper. In spite of the fact that the tongues
are very small, there is still energy associated with
gravitational acceleration of matter in the tongues and
the corresponding heating of the stellar surface. The hot spots
associated with gravitational acceleration are located in
the regions of highest density and occupy a much smaller area than
the spots shown in \fig{mbl-tiny-12}.
At the end of the $\widetilde{\mu}=0.1$ simulation run the
accretion rate increased, the disk reached the surface of the star and started
interacting with the star's surface through an equatorial belt
(see \S4.3).

We also performed exploratory simulations of accretion to stars with larger magnetospheres,
$\widetilde{\mu}= 0.5$ and observed that in many cases accretion through two or one ordered tongues dominates.
We varied the $\alpha$ parameter (which regulates the accretion rate)
and rotation rate of the star and obtained tongues and hot spots with
 different levels of order.
In some cases two identical, diametrically opposite streams dominate
accretion during the whole simulation run.
In other cases only one stream dominates while the other is weaker, and therefore only
one hot spot will determine the frequency observed in the light curve.
In yet another type of case, multiple tongues are observed (like in cases of large magnetospheres, e.g. KR08a; KR08b),
but one or two tongues are stronger than others, and can give rise to a QPO peak.
Below we include the $\widetilde{\mu}= 0.5$ case as an example of
accretion to slightly larger magnetospheres. This case links the very small magnetospheres
considered in this paper with the much larger ones considered in KR08b, where QPOs of similar origin
are observed at some sets of parameters.

%=========================================================

\subsection{Frequency analysis}

We performed frequency analysis for all the above cases.
First we perform what we call {\it spot-omega}
analysis, which we find to be the most informative for analysis of our simulations.
Namely, we plot the equatorial distribution of the emitted energy flux
at different moments of time (see \fig{swf-all}, left column). We obtain diagonal lines
which show that the spots have a definite order in their rotation along the star's equator.
The slopes of these lines are proportional to the angular velocity of the spots.
We obtained a number of almost parallel lines which reflect multiple rotations of a
single spot with approximately the same angular velocity. We performed this spot-omega analysis
for all cases. \fig{swf-all} shows the spot-omega diagrams for the same time interval.
One can see that at smaller $\widetilde{\mu}$, the lines are steeper because
the inner disk is closer to the star and the rotation of the streams/spots is faster.
Near the lines we show the rotation frequencies of the corresponding spots in units of
the stellar rotation frequency.
Note that in the case of the smallest magnetosphere
($\widetilde{\mu}=0.1$), the streams become very small and the spots associated with
the energy flux due to gravitational acceleration become weak. In this case
we plot the distribution of the density along the equator.
Closer to the end of this simulation run, the disk comes to the surface of the star,
and the streams (and ordered lines) disappear (see bottom left panel of \fig{swf-all}).
In the case of the largest magnetosphere, $\widetilde{\mu}=0.5$, the spot-omega diagram
shows a little less order because the magnetosphere is stronger and the rotation of the tongues is
less synchronized.

Next we performed wavelet analysis of the light-curves from hot spots. In the cases with
$\widetilde{\mu}=0.5, 0.2, 0.15$ and $0.12$, the light-curve is calculated with the suggestion that
all the gravitational and thermal energy of the matter falling onto the star is converted into
thermal energy, which is radiated isotropically. The
light from the surface of the star is integrated in the direction of the observer,
whose line of sight makes an angle of $\i=90^\circ$ with the rotational axis of the star (see RUKL04 and KR05 for details).
In the case with $\widetilde{\mu}=0.1$, it is expected that energy is released mainly due
to friction between the rapidly moving streams and the star.
Calculation of the radiation from such friction is beyond of the scope of this paper.
As a first approximation we suggest that the emitted flux is proportional
to the matter density at the star's surface.
\fig{swf-all} shows wavelet spectra for all these cases. These plots exclude early times, when
the streams have yet to reach the star's surface.
Since the wavelet transform
uses a window of width $\Delta t$ centered at time $t$ to calculate the time-dependent frequency spectrum
of the lightcurve from time $t_1$ to time $t_2$, it is necessary to have
$t_1+\Delta t/2 \lesssim t \lesssim t_2 - \Delta t/2$. This results in a portion
of the wavelet plot being cut off. The width $\Delta t$ is inversely
related to the frequency, so lower frequencies are more strongly
affected by this restriction. One can see that in each case several frequencies are observed.
Comparisons of the wavelet frequencies with the spot-omega frequencies (left panels)
 show that one of
the frequencies is approximately twice as high as the frequency of the single spots
shown in the spot-omega diagram,
and hence corresponds to the motion of two spots along the surface of the star.

We also performed Fourier analysis of the light-curves.
For this analysis we chose only the time intervals during which steady rotation
of the spots was observed.
In the case with $\widetilde{\mu}=0.1$, we also excluded the late times
after the spots disappear and the boundary layer forms. The Fourier spectra
show similar peaks as the wavelet. Between all these plots, the peak corresponding to rotation
of two spots is most important. Thus we get an approximate agreement between the high-frequency
QPOs obtained in all three methods of the frequency analysis.

There are also lower frequency peaks observed in both the wavelet and Fourier plots. These
frequencies may either reflect the shape of the spots, or be beat
frequencies \citep[see also][]{SmithLewisBonnell95, PsaltisEtAl98}.
We do not see the star's rotation frequency itself. However, it is possible that
our simulations are not long
enough for the wavelet analysis to capture possible QPOs associated
with the slowly rotating star. On the other hand, the spot-omega diagrams
do not show signs of stellar rotation, so it is possible that the corresponding
QPO is very weak.

Note that spiral density waves often form in the disk
(see Fig. \ref{mbl-tiny-12}, middle panels).
The density waves tend to be stationary in the
coordinate system rotating with a star, that is, they are generated
by the inclination of the dipole. The rotating tongues disrupt the inner
part of this pattern, but the external pattern approximately corotates
with the star (see \fig{mbl-tiny-12}). The figure also shows that
sample magnetic field lines starting out equidistantly from the star's surface
also accumulate in the density waves.
Note that these density waves are similar in shape to those
suggested by \citet{MillerLambPsaltis98}, except that
here the streams are guided by magnetic fields instead of being
produced by radiation drag.

%=========================================================

\subsection{Interaction through the boundary layer}

Closer to the end of the simulation run with the smallest magnetosphere
($\widetilde{\mu}=0.1$), the accretion rate
increased and the disk matter started interacting with the surface
of the star along a belt-like path (see \fig{mbl-KH-12}).
Later, the belt became wider, and an unusual pattern formed,
connected with some instability. This is probably the
Kelvin-Helmholtz instability which develops between the slowly rotating
stellar surface and the much more rapidly rotating disk, which is
expected to rotate at a frequency $\nu_{disk}\approx 8 \nu_*$ at the
surface of the star.

We should note that in all cases, when the modified tongue or a
boundary layer reaches the surface of the star, it spreads to higher
latitudes as it was predicted by \citet{InogamovSunyaev99} \citep[see
also][]{PiroBildsten04, FiskerBalsara05}. \fig{KH-lift} shows such spreading
and also a closer view of the
unstable boundary layer on the star's surface. During
such close contact between the disk and the star, energy is expected
to be released mainly from friction between the disk matter and the
stellar surface, like in the usual hydrodynamic boundary layer.
Compared to all the cases in the previous sections, no energy is
associated with the matter falling onto the star's surface.
Such boundary layers require special investigation. Note
that even in the case with no magnetosphere, a weak magnetic field
threads the disk, and magnetic field lines are wrapped in the inner parts
of the disk.

%%%%%%%%%%%%%%%%%%%%%%%% Fig 6 %%%%%%%%%%%%%%%%%%%%%%%%%%%%%%%

\figwide{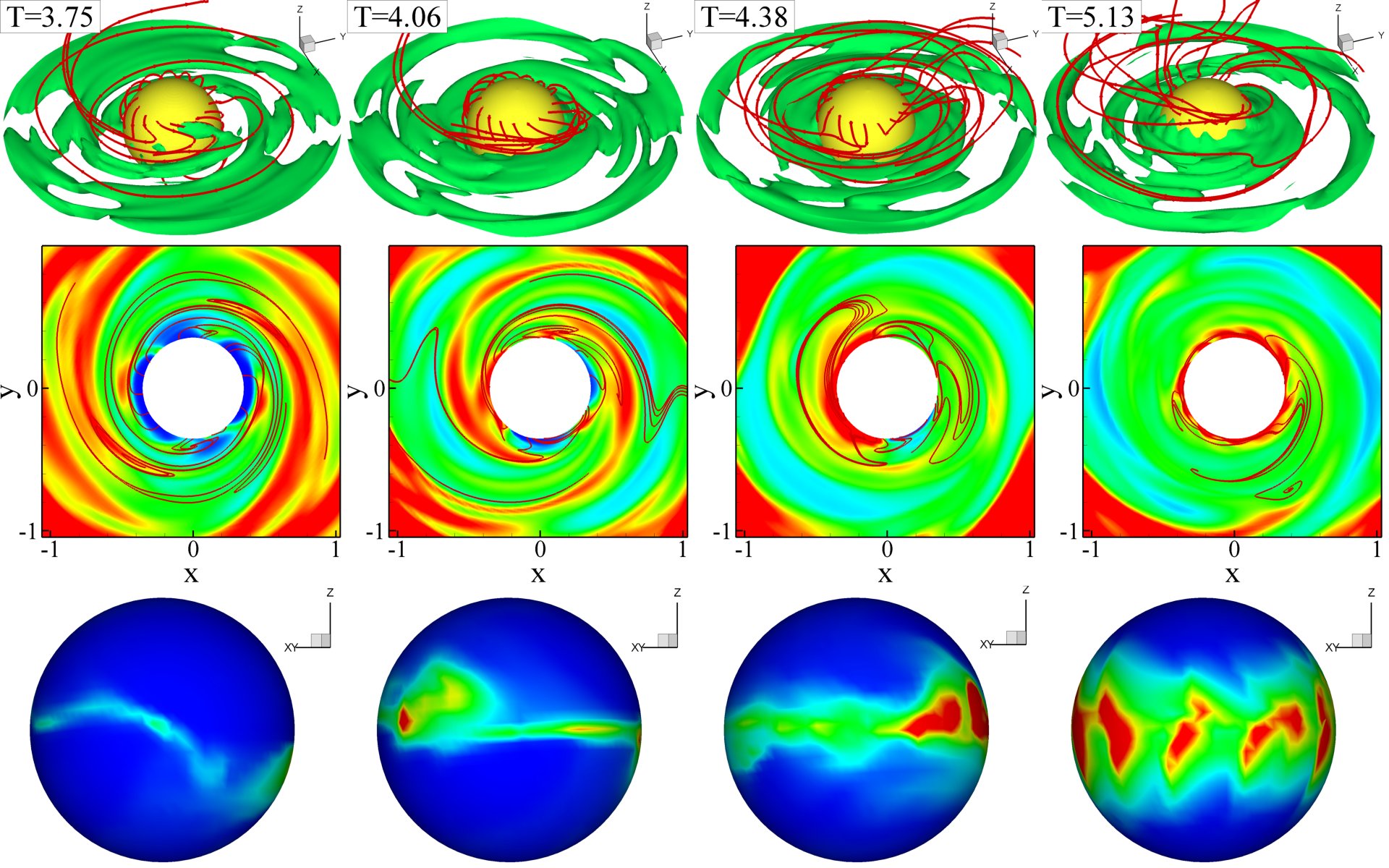}{mbl-KH-12}{Same plot as in \fig{mbl-tiny-12} but
closer to the end of the run when the disk approaches the star.
The surface shown in the top panels has a density of 0.43. In the bottom panels
the density varies between 2.6 (red) and 0.007 (blue).
The figures are shown in a coordinate system rotating with the star.
The time $T$ is measured in periods of Keplerian rotation at $r=1$.}

%%%%%%%%%%%%%%%%%%%%%%%% Fig 7 %%%%%%%%%%%%%%%%%%%%%%%%%%%%%%%

\fignar{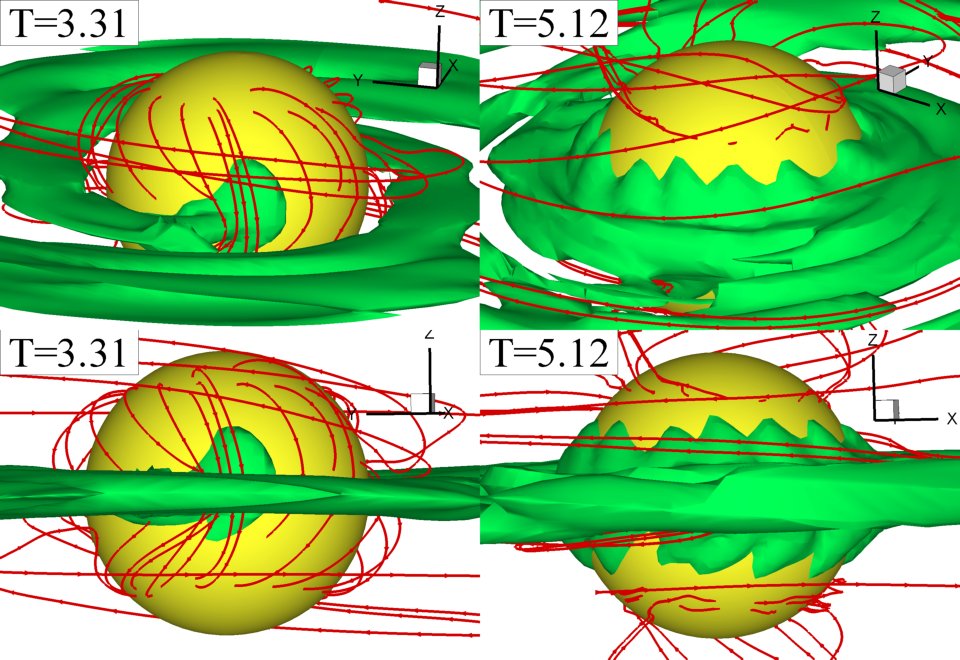}{KH-lift}{Enlarged view of two snapshots from \fig{mbl-KH-12},
showing a surface of constant
density (0.43 in dimensionless units) and sample magnetic field lines.
The top and bottom panels show different projections.
One can see that matter is lifted
along the surface of the star to larger latitudes. In the case when
the disk comes to the surface of the star, a new instability
appears at the disk-star boundary, which is probably of the
Kelvin-Helmholtz type.}

At later times, when
the disk comes closer to the star, the magnetic field lines trapped in
the equatorial region are pushed closer to the star or buried. Note
that at the same time the field lines above and below the disk are
not buried (see also \fig{KH-lift}). Possibly the process of field burial
by the thin disk through the boundary later should be reconsidered
taking into account the fact that a significant amount of magnetic flux
may inflate into the corona and stay there.
%=========================================================

\begin{table*}

\begin{tabular}{lllllll}
%______________________________________________________________________________________
\hline  & $\widetilde{\mu}=0.5$ & $\widetilde{\mu}=0.2$ & $\widetilde{\mu}=0.16$ & $\widetilde{\mu}=0.15$ & $\widetilde{\mu}=0.12$ & $\widetilde{\mu}=0.1$      \\
%__________________________________________________________________________________
\hline
%$\widetilde{\dot M}$          & 0.22   & 0.12   & 0.105  & 0.1  & 0.08     & 0.06                         \\
$\widetilde{\dot M}/\widetilde{\mu}^2$ & 0.88   & 3.0    & 4.1    & 4.4  & 5.5    & 6                      \\
$\dot M$ ($M_\odot$yr$^{-1}$) & $1.9\e{-9}$ & $6.6\e{-9}$ & $9.0\e{-9}$ & $9.7\e{-9}$ & $1.2\e{-8}$   & $1.3\e{-8}$    \\
$\nu_{1 spot}$ (Hz)         & 350   & 541 & 670  & 715 & 843  & 986                                        \\
$\nu_{2 spot}$ (Hz)        & 700  & 1082  & 1340  & 1430      & 1686  & 1972                       \\
%__________________________________________________________________________________
\hline
\end{tabular}

\caption{Values of the quantities described in \S 5.4 as a function of the magnetospheric size
parameter $\widetilde{\mu}$.} \label{tab:mudep}

\end{table*}

\subsection{Example: Application to Accreting Millisecond Pulsars}

Here we show a sample application of our simulations to
accreting millisecond pulsars.  We take a neutron star with mass
$M_*=1.4 M_\odot$, radius $R_*=10$ km and surface magnetic field
$B_*=3\e8$ G. The corresponding reference
values are in Table \ref{tab:refval}. We convert the dimensionless
results obtained in \S4.1
and \S4.2 into dimensional values.
The dimensionless stellar angular velocity is
$\widetilde{\omega_*}=0.354$.
The corresponding dimensional angular velocity is
$\omega_*=\widetilde{\omega_*}\omega_0=1002$ s$^{-1}$, dimensional period
is $P_*=2\pi/\omega_*= 6.3$ ms, and dimensional frequency is $\nu_*=159$
Hz. The time $T$ used in the figures has the reference value $P_0=2.2$ ms
which is the Keplerian rotation period at  $r=1$ ($3 R_*=30$ km).
The spot-omega diagrams in \fig{swf-all} (left panels) show the frequencies associated
with the rotation of
one spot for different values of $\widetilde{\mu}$.
In the case of the larger
magnetosphere ($\widetilde{\mu}=0.2$) the frequency is
$\nu_{1 spot}\approx 3.4 \nu_* =541$ Hz.  Since there are two antipodal hot
spots, we expect the observer to see twice that frequency,
$\nu_{2 spots}\approx 1082$ Hz. In the case with $\widetilde{\mu}=0.15$, the disk is closer to the star and
the frequencies are $\nu_{1 spot}\approx 4.5\nu_*\approx 715$ Hz and
$\nu_{2 spots} \approx 1430$ Hz. For $\widetilde{\mu}=0.12$ the frequencies are
$\nu_{1 spot}\approx 5.3\nu_*=843$ Hz and $\nu_{2 spots}\approx 1686$ Hz.
And for $\widetilde{\mu}=0.1$ the frequencies are $\nu_{1 spot}\approx 6.2\nu_*=986$ Hz
and $\nu_{2 spots}\approx 1972$ Hz. The frequencies are quite high because
the disk is close to the surface of the star. In simulations with a larger magnetosphere,
$\widetilde{\mu}=0.5$, we obtain lower frequencies, $\nu_{1 spot}\approx 2.2 \nu_* =350$ Hz and
$\nu_{2 spots}\approx 700$ Hz.

In application to neutron stars the formulae (1) can be re-written as:
$$
\dot M = 2.2\times10^{-9}\bigg(\frac{\widetilde{\dot M}}
{{\widetilde\mu}^2}\bigg) \bigg(\frac{B_*}{3\times10^{8} {\rm G}}\bigg)^2
$$
$$
\bigg(\frac{M_*}{2.8\times 10^{33}{\rm g}}\bigg)^{-\frac{1}{2}}
\bigg(\frac{R_*}{10^6 {\rm cm}}\bigg)^{\frac{5}{2}} ~ \frac{\rm{\dot M_{\odot}}}{\rm yr}~.
\eqno(2)
$$
One can see that the accretion rate depends on the dimensionless parameter $\widetilde{\dot M}/{\widetilde\mu}^2$.
Table \ref{tab:mudep} shows that
%the dimensionless accretion rate $\widetilde{\dot M}$ increases
%with the magnetospheric size parameter
%$\widetilde{\mu}$ (which probably reflects slightly enhanced braking in the disk connected with the disk-magnetosphere
%interaction). However, ${\widetilde{\mu}}^2$ varies more rapidly, and we see that
the ratio $\widetilde{\dot M}/{\widetilde\mu}^2$ decreases systematically with $\widetilde{\mu}$,
and so do the accretion rate
and the frequencies $\nu_{1 spot}$ and $\nu_{2 spots}$.
So, the considered above cases with different $\widetilde{\mu}$ correspond to
different accretion rates
where, depending on accretion rate, the frequency drifts between
$\nu_{2 spots}\approx 700$ Hz when the accretion rate is lower and
 the magnetosphere is larger, and
$\nu_{2 spots}\approx 1972$ Hz when the accretion rate is higher and
the magnetosphere is very small.
The ratio between these two frequencies is
2.8 and can be smaller or possibly larger depending on the variation of
the accretion rate, and is close to the observed drifts of the main
high-frequency QPO in, e.g., millisecond pulsars \citep{vanderKlis00}
and dwarf novae \citep{PretoriusWarnerWoudt06}.

Only if the two hotspots are antipodal and identical, do we expect $\nu_{1spot}$ to be
 absent from the power spectrum. This is true for the small magnetosphere
($\widetilde{\mu}\leq 0.2$) cases. In the large magnetosphere ($\widetilde{\mu}=0.5$) case, we see
that the spots are not identical; one spot is often much larger than the other,
and hence the frequency $\nu_{1spot}$ may dominate. If the magnetic field of the star
is not an ideal dipole field (that is, if it is a slightly misplaced dipole,
or a more complex field
\citep[e.g.,][]{LongRomanovaLovelace07,LongRomanovaLovelace08}
then the symmetry breaks and one spot will be always larger than the others, and
the lower frequency $\nu_{1spot}$ will dominate. Then we expected the frequency frift
between 350 Hz and 990 Hz.

Simulations of stars with larger magnetosphere sizes, $\widetilde{\mu}\gtrsim 0.5$,
usually show much more chaotic behavior in the
unstable regime (KR08a,b; RKL08). In spite of stochastic accretion in the unstable regime,
the spots on the surface of the star show a component which
rotates with the angular velocity of the inner disk (KR08b). This component analyzed through
rotation of spots (like in the left column of \fig{swf-all}) shows clear inclination of lines
associated with motion of different temporary spots
with angular velocity approximately equal to the angular velocity of the inner disk (KR08a).
This frequency component is weaker than the stochastic components associated with unstable accretion.
However in longer simulation runs this QPO peak associated with the inner disk frequency
may by amplified with time and may become significant, because the high-amplitude stochastic components
are relatively incoherent and their frequencies constantly drift.

Although we consider accretion to a very slowly rotating star, similar
MBLs and QPOs are expected for more rapidly rotating stars.

\subsection{Disk oscillations}

Note that some peaks observed
in the wavelets and Fourier spectra may be associated with the disk-star interaction.
An accretion disk can have different modes of oscillation.
These include  bending oscillations of the inner disk
driven by the star's rotating misaligned dipole field \citep[e.g.,][]{LaiZhang08} and
radial oscillations of the inner disk
\citep[e.g.,][]{AlparPsaltis08, LovelaceRomanova07,
 LovelaceTurnerRomanova09, ErkutEtAl08}.
The bending oscillations lead to the formation of an $m=1$ mode spiral wave
rotating with the frequency of the star.
In our simulations we do see traces of bending waves.
The radial oscillations can arise from
a linear Rossby wave instability (RWI) where
the angular  frequency of the  oscillations
(with mode number $m=1$) is less than the peak in the angular velocity in the
inner disk, $\Omega_{\rm max}$ \citep{LovelaceRomanova07, LovelaceTurnerRomanova09}.
This mode is radially trapped in a narrow
region inside the radius at which $\Omega_{\rm max}$ occurs.
The beat between this mode and the
stellar rotation frequency may lead to the coupled
twin peak QPOs observed in some
millisecond pulsars \citep[e.g.,][]{vanderKlis06}.

Here we derive from our simulations
the radius at which the disk angular velocity matches that of the spot.
\fig{rho-om-12} shows this analysis for cases with different $\widetilde{\mu}$.
First, we plot the radial dependence of the angular velocity in the disk
at sample times (see the top panels of the \fig{rho-om-12}).
  Then we take the angular velocity
of the spot from the spot-omega diagram (\fig{swf-all}, left panels) and find the point where
angular velocity of the spot equals
the angular velocity in the disk (see red dots).
   Next, we plot
the radial dependence of angular momentum (\fig{rho-om-12} middle panels)
and density (bottom panels) at same moment of time.
The circular line in the plots is the line on which the angular velocity  equals that
of the spot.
One can see that in all cases the rotational velocity of the spot corresponds to the inner edge of the disk.
It corresponds to a distance only slightly
larger than the Alfv\'en surface (red line), where the kinetic plasma parameter $\beta_1=(p+\rho v^2)/(B^2/8\pi)=1$.
Note that the magnetosphere is strongly non-axisymmetric, and therefore the $\Omega$ distribution in
the disk varies with time.

Different oscillation modes in the disk may influence the position or brightness variation
of the spots on the surface of the star. We do observe some frequencies in the power spectra of
the light-curves, but at present we are not sure that the observed frequencies reflect disk
oscillations. Future (longer) simulation runs may help reveal such a possibility. On the other hand,
disk oscillations can be investigated directly from the simulations. We plan to do this
in the future.

\fig{swf-all} shows that there are several peaks observed in the wavelet and Fourier spectra.
We know the origin of only
one of them, associated with rotation of the unstable tongues.
Other peaks may be associated with disk oscillations or with beat frequencies between
the disk and the star. However, at present our runs are not long enough to establish
the origin of these peaks.

%%%%%%%%%%%%%%%%%%%%%%%% Fig 8 %%%%%%%%%%%%%%%%%%%%%%%%%%%%%%%

\figwide{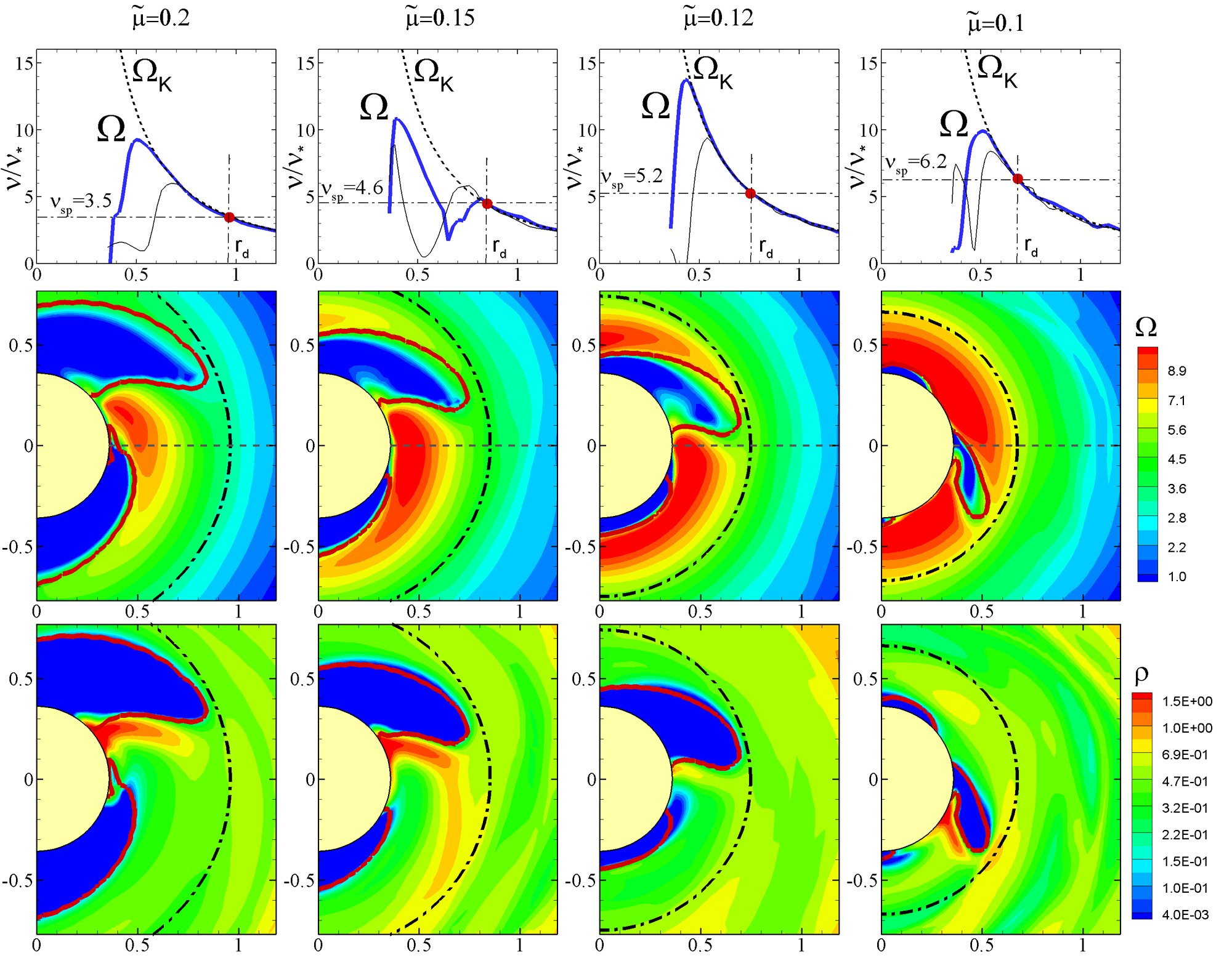}{rho-om-12}{{\it Top panels:} Radial dependence of the disk angular velocity for
the moments of time corresponding to the lower panels (thick blue line)
and for other moments of time (black line). The red spot marks the point where the angular velocity
matches that of the spots. {\it Middle panel:} Angular velocity distribution. The background
shows the angular velocity, while the foreground shows the kinetic plasma parameter $\beta_1=1$ (red thick line)
and the circle shows where the angular velocity equals that of the spot (black dash-dot line).
{\it Bottom panels:} Same, but for the density distribution.}
%=========================================================

\section{Conclusions}

We considered accretion to slowly rotating stars with small and tiny
magnetospheres in the stable and unstable regimes. We
conclude that:

(1) In the {\it stable regime}, a small magnetospheric cavity and tiny
funnel streams form, producing hot spots on the star's surface which
tend to be in a preferred position determined by
the dipole inclination (we had example runs for $\Theta=15^\circ$).
In this case the frequency of the star is expected to dominate.
However, the rapidly rotating
disk has a tendency to ``drag'' the funnel stream to faster rotation,
due to which parts of the hot spots often rotate much faster than the
star. At lower $\Theta$ this effect becomes even more significant, and the spots
may rotate faster than the star for a long time, leading to QPOs (see also RUKL04),
or slower than the star leading to matter accreting through a trailing funnel,
(e.g., Romanova et al. 2002) producing a lower-frequency QPO.

\smallskip

(2) In the {\it unstable regime} we observed that matter accretes through two ordered streams
which rotate with the angular velocity of the inner disk, that is, much faster than the star.
They hit the surface of the star forming two antipodal hot spots. Rotation of these spots
along the surface of the star leads to high-frequency QPOs. Such persistent
streams/spots have been observed at a variety of parameters and are seen to be quite long-lived.
Coherent rotation of this
kind has been observed for small magnetospheres.
For large magnetospheres we find that
the spots are much more chaotic in the sense that they form at different parts of the star
(RKL08; KR08a). In intermediate cases we observed that
one or two ordered streams may form, which are less ordered than in the cases
of small magnetospheres, but still may give QPO peaks (KR08b). On the other hand,
the accretion may be stochastic,
but one or two streams may be stronger than the others, and the hot spots
associated with these streams may lead to QPOs.

\smallskip

(3) Correlation is observed between the size of the magnetosphere and the QPO
frequency: the frequency is higher at smaller magnetospheric sizes. We expect
that secular variation of the accretion rate will lead to drifting of the QPO frequency.
Correlation between the frequency and the accretion rate has been observed in a number of
accreting millisecond pulsars \citep[e.g.,][]{vanderKlis00}.

\smallskip

(4) We were able to model a wide range of QPO frequencies. In application
to millisecond pulsars
the frequency associated with rotation of one spot  varies between
$\nu_{1spot}=350$ Hz and $990$ Hz.
In cases of small magnetospheres, the two spots are very similar in brightness,
and the expected frequencies are twice as high.
Only if the two spots are antipodal and identical, do we expect
$\nu_{1spot}$ to be absent from the power spectrum. This is true for the small
 magnetosphere ($\widetilde{\mu} \leq$ 0.2) cases.
 In the large magnetosphere ($\widetilde{\mu}$=0.5) case, we expect to see $\nu_{1spot}$.

\smallskip

The most striking result of this paper is the discovery of a
new regime of unstable accretion which shows clear high-frequency
QPO peaks.

%#########################################################

\subsubsection*{Acknowledgments}
The authors thank R.V.E. Lovelace for discussions and the referee M.A. Alpar
for comments and suggestions which helped improve the paper.
The authors were supported in part by
NASA grant NNX08AH25G and by NSF grants AST-0607135 and AST-0807129.
We are thankful to NASA for using NASA High Performance Facilities.

%#########################################################

\bibliography{ms-new}

\end{document}